\documentclass[10pt]{article}
\usepackage{amsfonts,latexsym}

\catcode`\@=11
\def\Let@{\relax\iffalse{\fi\let\\=\cr\iffalse}\fi}
\def\vspace@{\def\vspace##1{\crcr\noalign{\vskip##1\relax}}}
\def\multilimits@{\bgroup\vspace@\Let@
 \baselineskip\fontdimen10 \scriptfont\tw@
 \advance\baselineskip\fontdimen12 \scriptfont\tw@
 \lineskip\thr@@\fontdimen8 \scriptfont\thr@@
 \lineskiplimit\lineskip
 \vbox\bgroup\ialign\bgroup\hfil$\m@th\scriptstyle{##}$\hfil\crcr}
\def\Sb{_\multilimits@}
\def\endSb{\crcr\egroup\egroup\egroup}
\def\Sp{^\multilimits@}

\catcode`\@=11
\def\lesssim{\mathrel{\mathpalette\vereq<}}
\def\vereq#1#2{\lower3pt\vbox{\baselineskip1.5pt \lineskip1.5pt
\ialign{$\m@th#1\hfill##\hfil$\crcr#2\crcr\sim\crcr}}}

\newcommand{\be}[1]{\begin{equation}\label{#1}}
\newcommand{\ee}{\end{equation}}
\newcommand{\ba}[1]{\begin{eqnarray}\label{#1}}
\newcommand{\ea}{\end{eqnarray}}
\newcommand{\rf}[1]{(\ref{#1})}
\newcommand{\nn}{\nonumber}
\newcommand{\bmatrix}[1]{\left( \begin{array}{#1}}
\newcommand{\ematrix}{\end{array}\right)}

\newlength{\indentedwidth}
\newdimen\mathindent
\indentedwidth=\mathindent

\newenvironment{indented}{\begin{indented}}{\end{indented}}
\newenvironment{varindent}[1]{\begin{varindent}{#1}}{\end{varindent}}
\def\indented{\list{}{\itemsep=0\p@\labelsep=0\p@\itemindent=0\p@
   \labelwidth=0\p@\leftmargin=\mathindent\topsep=0\p@\partopsep=0\p@
   \parsep=0\p@\listparindent=15\p@}\footnotesize\rm}

\def\varindent#1{\setlength{\varind}{#1}%
   \list{}{\itemsep=0\p@\labelsep=0\p@\itemindent=0\p@
   \labelwidth=0\p@\leftmargin=\varind\topsep=0\p@\partopsep=0\p@
   \parsep=0\p@\listparindent=15\p@}\footnotesize\rm}

\begin{document}

\author{U. G\"unther\dag\footnote{e-mail: u.guenther@htw-zittau.de} \ and
A. Zhuk\ddag\footnote{permanent address:
Department of Physics, University of Odessa, 
2 Petra Velikogo St., Odessa 270100, Ukraine; \newline \indent \ \ \
e-mail: zhuk@paco.odessa.ua} \\[2ex]
\dag Gravitationsprojekt, Mathematische Physik I,\\
Institut f\"ur Mathematik,
Universit\"at Potsdam,\\
Am Neuen Palais 10, PF 601553, D-14415 Potsdam, Germany\\[1ex]
\ddag Centro de F\'{\i}sica ``Miguel Catal\'{a}n'',\\
Instituto de Matem\'{a}ticas y F\'{\i}sica Fundamental,\\
Consejo Superior de Investigaciones Cient\'{\i}ficas,\\ 
Serrano 121, 28006
Madrid, Spain}
\title{Gravitational Excitons as Dark Matter
\thanks{Report given
at the {\it Conference on Cosmology and Particle
Physics, CAPP 2000}, 
Verbier, Switzerland,  July 17-28, 2000.}
}

\date{01.11.2000}

\maketitle

\begin{abstract}
In earlier work it was pointed out that for warped product spacetimes
the conformal  (geometrical moduli) excitations
of the internal compactified factor spaces should be observable
as massive scalar fields 
in the external spacetime. Here we show that these scalar fields
(gravitational excitons) describe
weakly interacting particles and can be considered as dark matter component.
Masses of the gravexcitons are
defined by the form of the effective potential of the theory 
and the stabilization scales
of the internal space. This implies that different
stabilization scales result in different types of DM. An essential role
is played by the effective potential.
On the one hand, its minima fix possible stabilization scales of the internal
spaces; on the other hand, they provide possible values for the
effective cosmological constant.
\end{abstract}

\bigskip

\hspace*{0.950cm} PACS number(s): 04.50.+h, 98.80.Hw
%

\section{Introduction}

\setcounter{equation}{0}

Modern observations lead to the conclusion that the cosmological
constant $\Lambda$ contributes about 0.7 to the total $\Omega$, i.e.,
the cosmological constant is non-zero and positive.
Matter contributes another 0.3 to $\Omega$ and most
of it consists of Dark  Matter (DM). This implies that the Universe
is spatially flat to a good approximation.
The claim that the Universe is spatially flat
is extremely strong: it results from the position of
the first Doppler/Sakharov peak in the CMB anisotropy.
This position is very well measured, notably, by
the Boomerang experiment. One of the most important problems
in modern cosmology is to construct a viable theoretical model
which naturally explains the observed value of the cosmological constant
and the DM contribution to $\Omega$.

In the present paper we show with the help of a simple toy model that
extra dimensions could give a good background
for the resolution of these problems. More concrete, conformal
excitations of internal spaces (geometrical moduli excitations)
will propagate in our Universe
as massive scalar fields (gravexcitons) and play the role of DM.
Minima of their potential energy are treated as the $\Lambda$ term.

\section{Noninteracting Gravitational Excitons}

\setcounter{equation}{0}

We consider a cosmological toy model with metric
\begin{equation}
\label{2.1}g=g^{(0)}+\sum_{i=1}^ne^{2\beta ^i(x)}g^{(i)},
\end{equation}
which is defined on a manifold with warped product topology
\begin{equation}
\label{2.2}M=M_0\times M_1\times \dots \times M_n\, .
\end{equation}
Here $x$ denotes some coordinates of
the $D_0 =(d_0+1)-$dimensional manifold $M_0$ and
\begin{equation}
\label{2.3}g^{(0)}=g_{\mu \nu }^{(0)}(x)dx^\mu \otimes dx^\nu .
\end{equation}
Let manifolds $M_i$ be $d_i-$dimensional Einstein spaces with metric $%
g^{(i)},$  i.e.
\begin{equation}
\label{2.4}R_{mn}\left[ g^{(i)}\right] =\lambda ^ig_{mn}^{(i)},\qquad
m,n=1,\ldots ,d_i
\quad \mbox{and} \quad
R\left[ g^{(i)}\right] =\lambda ^id_i\equiv R_i.
\end{equation}
For constant curvature spaces the parameters $\lambda ^i$ are
normalized as $\lambda ^i=k_i(d_i-1)$ with $k_i=\pm 1,0$. Later on we shall
not specify the structure of the spaces $M_i$. We require only $M_i$ to be
compact spaces with arbitrary sign of curvature.

With total dimension $D= D_0 +\sum_{i=1}^nd_i$,\, $\kappa_D ^2$ a
$D-$dimensional
gravitational constant, and $\Lambda $ - a $D-$dimensional cosmological
constant, we consider an action of the form
\begin{equation}
\label{2.6}S=\frac 1{2\kappa_D ^2}\int\limits_Md^Dx\sqrt{|g|}\left\{
R[g]-2\Lambda \right\} + S_m\, .
\end{equation}
 $S_m$ is a not specified action-term of matter fields. 
To illustrate the  origin of
 gravexcitons it is sufficient to consider a pure geometrical model with
$S_m \equiv 0$. Generalizations to models with included matter are obvious, 
and for different types of matter 
they can be found in our papers \cite{GZ,GZ(PRD2)}.

Let $\beta^i_0$ be the scale of compactification of the internal spaces
at present time. Instead of $\beta^i$ it is convenient to
introduce the shifted quantity: $\tilde \beta^i = \beta^i - \beta^i_0$.

Then, after dimensional reduction and conformal transformation
\begin{equation}
\label{2.12} g_{\mu \nu }^{(0)}= \Omega^2 \tilde g_{\mu \nu
}^{(0)} := {\left( \prod_{i=1}^ne^{d_i\tilde \beta ^i}\right) }
^{\frac{-2}{D_0-2}}
\tilde g_{\mu \nu }^{(0)}
\end{equation}
action \rf{2.6} reads
\begin{equation}
\label{2.13}S=\frac 1{2\kappa _0^2}\int\limits_{M_0}d^{D_0}x\sqrt{|\tilde
g^{(0)}|}\left\{ \tilde R\left[ \tilde g^{(0)}\right] -\bar G_{ij}
\tilde g^{(0)\mu \nu }\partial _\mu \tilde \beta ^i\,
\partial _\nu \tilde \beta ^j - 2U_{eff}\right\} ,
\end{equation}
where $\tilde R_i := R_i e^{-2\beta^i_0}$,
$\bar G_{ij}=d_i\delta _{ij}+\frac 1{D_0-2}d_id_j$
is the midisuperspace metric and
$\kappa^2_0 := \kappa^2_D/V_{D^{\prime }}$
denotes the $D_0-$dimensional gravitational constant. 
($V_{D^{\prime }}$ is the
total volume of the internal space.)
If we take the TeV scale $M_{TeV}\sim 1$ TeV and the Planck scale 
$M_{Pl}\sim 1.22\times 10^{19}$ GeV
as fundamental scales for the $D-$dimensional total spacetime 
and the 4-dimensional large scale spacetime,
respectively:
$\kappa^2_D = 8\pi/M^{2+ D^{\prime}}_{TeV}\,$, 
$\kappa^2_0 = 8\pi/M^2_{Pl}\,$
then we reproduce the well known relation \cite{sub-mill}:
~$M_{Pl}^2 = V_{D^{\prime}} M_{TeV}^{(2+D^{\prime})}\,$.
Thus, the compactification scale of the internal space
is fixed and of order
\be{1.2}
a \sim V_{D^{\prime}}^{1/D^{\prime}}
\sim 10^{32/D^{\prime}-17} \mbox{cm}\, .
\ee
The effective potential in (\ref{2.13}) reads
\begin{equation}
\label{2.14}U_{eff}[\tilde \beta ] =
{\left( \prod_{i=1}^ne^{d_i\tilde \beta ^i}\right) }^{-\frac
2{D_0-2}}\left[ -\frac 12\sum_{i=1}^n\tilde R_ie^{-2\tilde \beta ^i}+\Lambda
\right]\, .
\end{equation}

With the help of a regular coordinate transformation
$\varphi =Q\tilde\beta ,\quad \tilde\beta =Q^{-1}\varphi$
midisuperspace metric (target
space metric) $\bar G$ can be transformed to a pure Euclidean form:
$\bar G_{ij}d\tilde\beta ^i\otimes d\tilde\beta ^j
=\sigma _{ij}d\varphi ^i\otimes
d\varphi ^j=\sum_{i=1}^nd\varphi ^i\otimes d\varphi ^i,\quad
\sigma ={\rm diag\ }(+1,+1,\ldots ,+1)$ (see e.g. \cite{GZ}).

Clearly, a stabilization of the internal spaces can be achieved
if the effective potential $U_{eff}$ has a minimum with respect to fields
$\tilde \beta^i$ (or fields $\varphi^i$). In general it is possible
for the potential $U_{eff}$
to have more than one extremum.
For the pure geometrical toy model under consideration
 we can get only one extremum, 
which corresponds to  $\tilde \beta^i = 0$.
For the masses of the normal mode excitations
of the internal spaces (gravitational excitons)
around the minimum position we obtain:
\be{2.22}
m_1^2 = \dots =m_n^2 = -\frac{4\Lambda_{eff}}{D_0-2}=-2\frac{\tilde R_k}
{d_k} > 0\, ,
\ee
where
\be{2.21}
\Lambda_{eff} :=\left. U_{eff}\vphantom{\int} \right|_
{\tilde \beta^i =0}\,
\ee
plays the role of an effective cosmological constant in the external
spacetime.
These equations show that for our specific model
a global minimum can only exist
in the case of compact internal spaces with negative curvature
$R_k <0\; (k=1,\dots ,n)$.
The effective cosmological constant is then
also negative: $\Lambda_{eff} <0$.
Models which include matter can have minima for internal spaces of positive
curvature, and the effective cosmological constant in this case is
usually positive.

For small fluctuations of the normal modes in the vicinity
of the minima of the effective potential, action \rf{2.13} reads
\begin{eqnarray}\label{2.23}
S & = & \frac{1}{2\kappa _0^2}\int \limits_{M_0}d^{D_0}x \sqrt
{|\tilde g^{(0)}|}\left\{\tilde R\left[\tilde g^{(0)}\right] - 2\Lambda
_{eff}\right\} - \\
\ & - & \frac{1}{2}\int \limits_{M_0}d^{D_0}x \sqrt
{|\tilde g^{(0)}|}\left\{\sum_{i=1}^n \left( \tilde g^{(0)\mu \nu}
\psi ^i_{,\mu}\psi^i_{,\nu} + m_i^2\psi ^i\psi ^i\right)
\right\}\, . \nn
\end{eqnarray}
(For convenience we use here the normalizations:
$\kappa^{-1}_0 \tilde \beta \rightarrow \tilde \beta$.)
Thus, conformal excitations of the metric of the internal spaces behave as
massive scalar fields developing on the background of the external
spacetime. In analogy with excitons in solid state physics where they are
excitations of the electronic subsystem of a crystal, we called
the excitations of the
subsystem of internal spaces  gravitational excitons
\cite{GZ}. Later, since \cite{sub-mill} these
particles are also known as radions.

{}From eq. (\ref{2.22}) follows that
\be{2.30}
|\Lambda_{eff}| \sim m^2_i \sim a^{-2}_{(0)i}\, ,
\ee
where $a_{(0)i} = \exp{\beta^i_0}$ are the scale factors
of stabilized internal spaces.

The calculations above were
performed in a model with  the TeV scale $M_{TeV}$
as  fundamental scale of the $D-$dimensional
theory (see eq. (\ref{1.2})).
Clearly, it is also  possible to choose the Planck scale
as the fundamental scale.

For this purpose, we will not fix the compactification scale
of the internal
spaces at their present time values. We consider them as free parameters of
the model and demand only that $L_{Pl} < a_{(0)i} = e^{\beta^i_0} <
L_{F}\sim 10^{-17}$cm. So, we do not transform
$\beta^i$ to $\tilde \beta^i$.
In this case, $\kappa^2_D \sim M^{-(2+D^{\prime})}_{Pl}$, so that
the Planck scale
becomes the fundamental scale of the $D-$dimensional
theory. In this approach eqs. \rf{2.12}, \rf{2.13} and \rf{2.14}
preserve their form, with only substitutions $\tilde \beta \longrightarrow
\beta$ and $\tilde R_i \longrightarrow R_i$. The Einstein frame
metrics of the external spacetime in both approaches are equivalent to each
other up to a numerical prefactor:
\be{c.4a}
\left. \tilde g^{(0)}_{\mu \nu}\right|_{TeV}
=v_0^{-2/(D_0-2)}\left.\tilde g^{(0)}_{\mu \nu}\right|_{Pl}\, ,
\ee
where $v_0 = \prod\nolimits_{i=1}^n \exp{(d_1\beta^i_0)}$. Obviously,
the same rescaling takes place for the
masses squared of the gravitational excitons, and the effective
cosmological constant: $m^2_i \longrightarrow \left( v_0 \right)^
{-2/(D_0-2)} m_i^2$ and
$\Lambda_{eff} \longrightarrow \left( v_0 \right)^
{-2/(D_0-2)} \Lambda_{eff}$. Thus, in the latter approach we get instead of
(\ref{2.30}) the relation:
\be{c.4}
|\Lambda_{eff}| \sim m_i^2 \sim
\left(a_{(0)i}\right)^{-(D-2)}\, ,
\ee
where we set $D_0=4$.
This expression shows that due to the power $(2-D)$
the effective cosmological constant and the masses of the gravitational
excitons can be very far from planckian values, even for scales of
compactification of the internal spaces close to the Planck length.

Let us return to the comparison of the TeV scale and the Planck scale
approaches. E.g., within the TeV scale approach for $6\le D <
\infty$ the internal space scale factors, gravexciton masses and effective
cosmological constant run, correspondingly, as: $10^{-1}$cm $\le a_{(0)i} <
10^{-17}$cm, $\quad 10^{-4}$eV $\le m_i <1$TeV and $10^{-64}\Lambda_{Pl}
\le |\Lambda_{eff}| \le 10^{-32}\Lambda_{Pl}$. For this approach
the scale factors of the internal spaces are defined by eq. \rf{1.2}, due
to the requirement that the $D-$dimensional gravitational constant
is of order of the TeV scale.
In the Planck scale approach this condition is absent, and $a_{(0)i}$ are
free parameters. Let us take, e.g., $a_{(0)i} \sim 10^{-18}$cm. Then,
within the Planck scale approach for $6 \le D \le 10$
the gravexciton masses and the effective
cosmological constant run correspondingly as:
$10^{-2}$eV $\le m_i \le 10^{-32}$eV and $10^{-60}\Lambda_{Pl}
\le |\Lambda_{eff}| \le 10^{-120}\Lambda_{Pl}$.

These estimates show that for the TeV scale  approach
the effective cosmological constant is much greater than the present
day observable limit
$\Lambda \le 10^{-122}\Lambda_{Pl} \sim 10^{-57}{\mbox cm}^{-2}$
(for our model
$\left. |\Lambda_{eff}|\right|_{TeV}\ge 10^2 \, \mbox{\rm cm}^{-2}$),
whereas in the
Planck scale approach we can satisfy this limit
even for very small compactification scales.
For example, if we demand in accordance with observations $|\Lambda_{eff}|
\sim 10^{-122}\Lambda_{Pl}$ then eq. (\ref{c.4}) gives a
compactification  scale  $a_{(0)1} \sim 10^{122/(D-2)}L_{PL}$.
Thus, $a_{(0)1} \sim 10^{15}L_{Pl} \sim 10^{-18}{\mbox cm}$ for $D=10$ and
$a_{(0)1} \sim 10^{5}L_{Pl} \sim 10^{-28}{\mbox cm}$ for $D=26$, which
is not in contradiction to observations because for this approach the
compactification scales should be
$a_{(0)1} \le 10^{-17}{\mbox cm}$. Assuming an estimate
$\Lambda_{eff}\sim 10^{-122}L_{Pl}$,
we automatically get from eq. (\ref{c.4}) the value
of the gravitational exciton mass: $m_1 \sim 10^{-61}M_{Pl} \sim
10^{-33}eV \sim 10^{-66}{\mbox g}$, which is extremely light. Nevertheless
such light particles are not in contradiction with observations,
because these particles do not overclose the Universe \cite{GZ(PRD2)}.

\section{Interacting Gravitational Excitons}

\setcounter{equation}{0}
By definition, Dark Matter (DM) consists of particles interacting with usual
matter mainly via gravitational forces.
Thus, to define gravexcitons as DM we should show that their interaction
with usual matter is very weak. Here, we do this with respect to their
interaction with electromagnetic (e.m.) fields. As we saw above, gravexitons
are neutral particles and for this reason cannot interact with e.m. fields via
 current. 
Their interaction with e.m. fields originates in their multidimensional
nature, and can be easily understood
if we consider a zero-mode approximation of the e.m. fields 
in the multidimensional spacetime.
Due to the multidimensional determinant of the metric 
(which depends on the scale factors of the internal
space), the dimensionally reduced action $S_m$ will contain
nonlinear terms describing the interaction between matter 
and gravexcitons\footnote{In the case of matter 
located on a 4-dimensional brane (which is
simulated by  delta-function-fixing of the brane position 
in the multidimensional
space) such interaction terms exist also due to the time dependence of the
 scale factors of the internal space.}.
In the simplest case of one internal space $(n=1)$,  
the gravexciton-photon interaction is in lowest order approximation
described by the term \cite{GSZ}:
\be{3.1}
2\sqrt{\frac{d_1}{(D_0-2)(D-2)}}\kappa_0 \psi F_{\mu \nu}
F^{\mu \nu}\, .
\ee
A corresponding first order diagram of this interaction describes the decay
of gravexcitons into photons: $\psi \longrightarrow 2\gamma$.
The probability of this decay is easily  estimated as
\be{3a}
\Gamma = \frac{2d_1}{d_1+2} \frac{m^3}{M^2_{Pl}} = \frac{2d_1}{d_1+2}
\left(\frac{m}{M_{Pl}}\right)^3\frac{1}{T_{Pl}} \ ,
\ee
which results in a life-time of the gravitational
excitons with respect to this decay
\be{4a}
\tau = \frac{1}{\Gamma} =\frac{d_1+2}{2d_1}
\left(\frac{M_{Pl}}{m}\right)^3 T_{Pl}
\, .
\ee
These relations show that gravexcitons with masses
$m \le 10^{-21} M_{Pl} \sim 10^{-2}\mbox{GeV} \sim 20 m_e$ (where
$m_e$ is the electron mass) have a life-time $\tau \ge 10^{19}\mbox{sec} >
t_{univ} \sim 10^{18}$ sec, which is greater than the age of the Universe.
Thus, they are stable particles with respect to this process.
In other words, such gravexcitons interact weakly (Planck scale suppressed)
with e.m. fields and can be considered as DM.
Similar estimates hold for the interaction of gravexcitons
with other types of matter. The type of the DM
depends on the DM particle masses. It is hot for $m_{DM} \le 50 - 100$eV,
warm for $ 100\mbox{eV} \le m_{DM} \le 10$KeV and cold for $m_{DM} \ge
10 - 50$KeV. Gravexciton masses are closely related with the compactification
scales of the internal
space  (see (\ref{2.30}) and (\ref{c.4})). As shown
in the previous section, gravexcitons may be hot DM as well as cold DM, 
depending
on $a_{(0)i}$. However, gravexcitons with masses \cite{GZ(PRD2)}
\be{c.10}
m \ge 10^{-56} M_{Pl} \left(\frac{M_{Pl}}{\varphi_{in}}
\right)^4
\ee
overclose the Universe, i.e. their energy density at the present time
is greater than the critical energy density of the Universe.
Usually, it is assumed that the amplitude of initial gravexciton
oscillations
$\varphi_{in} \sim O (M_{Pl})$
can be considerably less than $M_{Pl}$
(although it depends
on the form of $U_{eff}$). If we
assume $\varphi_{in} \sim O(M_{Pl})$ then excitons with masses $m
\lesssim 10^{-28}{\mbox eV}$ will not overclose the Universe.
Thus, gravexcitons are either very hot DM with masses $m \le 10^{-28}$eV
(and with negligible contribution to the total amount of Dark Matter)
or the amplitude of initial oscillations is $\varphi_{in} \ll M_{Pl}$
and gravexcitons may be cold DM.


\begin{thebibliography}{99}

\bibitem{GZ}
U. G\"unther and A. Zhuk, Phys. Rev. D56, (1997), 6391 - 6402,
\, gr-qc/9706050;\,
%
Stable compactification and
gravitational excitons from extra di\-men\-sions, (Proc. Workshop
{\it Modern
Modified Theories of Gravitation and Cosmology}, Beer Sheva, Israel, June
29 - 30, 1997), Hadronic Journal 21, (1998),  279 - 318, \,
gr-qc/9710086;\,
Class. Quant. Grav. 15,
(1998), 2025 - 2035, \, gr-qc/9804018;
U. G\"unther, S. Kriskiv and A. Zhuk, Gravitation
and Cosmology 4, (1998), 1 - 16, \,  gr-qc/9801013.

\bibitem{GZ(PRD2)}
U. G\"unther and A. Zhuk, Phys. Rev. D61:124001, (2000),
\, hep-ph/0002009.

\bibitem{sub-mill}
N. Arkani-Hamed, S. Dimopoulos and J. March-Russell, {\it Stabilization
of sub-millimetre dimensions: the new guise of the hierarchy
problem},\, hep-th/9809124;
N. Arkani-Hamed, S. Dimopoulos, N. Kaloper and J. March-Russell,
Nucl. Phys. B 567, (2000), 189 - 228,\, hep-ph/9903224.

\bibitem{GSZ}
U. G\"unther and A. Zhuk,
Interacting gravitational excitons and observable effects from
extra dimensions, (Proc. Memorial International Conference (GMIC'99)
{\ The Universe of Gamov: Original Ideas in Astrophysics and Cosmology},
Odessa, Ukraine, August 16 -22, 1999), Odessa Astronomical Publications
12, (1999) 37 - 47, \, http://oap12.webjump.com ;\,
U. G\"unther, A. Starobinsky and A. Zhuk, {\it Interacting gravitational
excitons from extra dimensions}, \, (in preparation).
%
\end{thebibliography}
\end{document}